\documentstyle[11pt, aaspp4]{article}

\newcommand{\gtsim}{\lower.5ex\hbox{$\; \buildrel > \over \sim \;$}}
\newcommand{\ltsim}{\lower.5ex\hbox{$\; \buildrel < \over \sim \;$}}

\newcommand{\beq}{\begin{equation}}
\newcommand{\eeq}{\end{equation}}

\begin{document}
 

%
%
\title{Large Scale Fluctuations in the X-Ray Background}

\author{Marie Treyer$^1$, Caleb Scharf$^2$, Ofer Lahav$^3$,
Keith Jahoda$^4$, Elihu Boldt$^4$, Tsvi Piran$^5$}

\affil{$^1$ Astrophysikalisches Institut Potsdam, 
An der Sternwarte 16, 14842 Potsdam, Germany, mtreyer@aip.de}

\affil{$^2$ Space Telescope Science Institute, 3700 San Martin Drive, 
Baltimore, MD 21218,  scharf@stsci.edu}

\affil{$^3$ Institute of Astronomy, Madingley Road, 
Cambridge CB3 OHA, U.K.; lahav@ast.cam.ac.uk}

\affil{$^4$ Laboratory for High Energy Astrophysics, 
NASA/GSFC, Greenbelt, MD 20771,
keith@pcasrv2.gsfc.nasa.gov, boldt@lheavx.gsfc.nasa.gov}

\affil{$^5$ Racah Institute of Physics, The Hebrew University, 
Jerusalem 91904, Israel, tsvi@shemesh.fiz.huji.ac.il}

\begin{abstract} 
We present an attempt to measure the large angular scale 
fluctuations in the X-Ray Background (XRB) from the HEAO1-A2 
data, expressed in terms of spherical harmonics. We model the
harmonic coefficients assuming a power spectrum and
an epoch-dependent bias parameter, $b_x(z)$, and using a 
phenomenological scenario describing the evolution of the X-ray 
sources.
From the few low-order multipoles detected above 
shot noise, we estimate the power-spectrum normalization on scales
intermediate between those explored by local galaxy 
redshift surveys ($\sim 100 h^{-1} $ Mpc) and those probed by the
COBE Cosmic Microwave Background (CMB) measurements 
($\sim 1000 h^{-1} $ Mpc).
We find that the HEAO1 harmonics are consistent with present epoch 
rms fluctuations of the X-ray sources   
$b_x(0) \sigma_8 \sim 1-2$ in 8 $h^{-1}$ Mpc  spheres.
Therefore the observed fluctuations in the XRB
are roughly as expected from interpolating between the
local galaxy surveys and the COBE CMB experiment.
We predict that an X-ray all-sky surface brightness
survey resolving sources a factor of 10 fainter than HEAO1, 
may reveal fluctuations to significantly larger
scales and therefore more strongly constrain 
the large scale structure of the Universe 
on scales of hundreds of Mpcs.
\end{abstract} 


\section{Introduction} 

Although discovered before the Cosmic Microwave Background (CMB), 
the origin of the hard X-Ray Background (XRB) is still not fully 
understood.
At energies below 2 keV, the XRB has now been almost entirely 
resolved into discrete sources. Most of these are AGN's but other
types of sources (e.g. clusters, narrow emission line galaxies)  
may also contribute a significant fraction of the total flux
(Hasinger et al.~1998, McHardy et al.~1998). 
The X-ray spectra of these sources are, for the most part, 
too soft to account for the shape and total intensity of the XRB
above 2 keV, but absorption can be invoked to remedy this problem
and allow the full energy range of the XRB to be nicely 
fitted by the right mixture of absorbed and unabsorbed AGN's 
(Comastri et al.~1995). Such models have their limitations,
but even if the detailed nature of the hard X-ray 
sources, and their relation to the soft ones, are not 
fully understood yet, it is now well established 
that the dominant part of the hard XRB also arises from 
the integrated emission of discrete sources.
The alternative hypothesis of a cosmic hot gas 
origin was ruled out by the observation of 
the undistorted CMB spectrum (see Fabian \& Barcons 
1992 for review).

In order to account for the total flux of the XRB 
(local X-ray sources only produce a very small fraction 
of it), X-ray sources must be found throughout 
a large enough volume of the universe. This 
makes them convenient tracers of the mass distribution
on scales intermediate between those probed by COBE in the CMB 
($\sim 1000 $ Mpc), and those probed by optical and IRAS 
redshift surveys ($\sim 100 $ Mpc). 
The CMB fluctuations originate from redshift $z \sim 1000$
and are due to the Sachs-Wolfe effect on scales larger than 
a few degrees. On the other hand, the fluctuations in the XRB
are due to fluctuations in the space density of
X-ray sources which are likely to be distributed 
at $z \sim 1-5$. In terms of level 
of anisotropy, the XRB is also intermediate between the CMB 
fluctuations ($\sim 10^{-5}$ on angular scales of
degrees) and the galaxy density fluctuations (of the order 
of unity on scale of 8 $h^{-1}$ Mpc). 

In this paper, we attempt a comparison between the hard band (2-10 keV) 
XRB fluctuations seen in the HEAO1-A2 data and a range 
of models. We measure the fluctuations in terms of
spherical harmonic coefficients, and make predictions
for the ensemble average of these coefficients using
a formalism presented by Lahav, Piran \& Treyer (1997) 
(hereafter LPT97). 
For related approaches to measurements 
of the XRB fluctuations see Boughn, Crittenden \& Turok (1998) 
and Carrera, Barcons \& Fabian (1997) and references therein.
The data analysis and the theoretical formalism are described 
in Sections 2 and 3 respectively. Measurements and models are 
compared in Section 4. We present our conclusions in Section 5.
For simplicity, we shall assume an Einstein-de Sitter
world geometry ($\Omega=1$, $\Lambda=0$). 
We write the Hubble constant as $H_0=100~h~{\rm km/s/Mpc}$. 

\section{The HEAO1-A2 Data Analysis}

Details of the HEAO1-A2 data analysis will be described in 
a complementary paper (Scharf et al., in preparation). 
This section summarizes the procedure. 

We use the A2 counts from the 
6 months following day 322 of 1977 in
the all-sky survey (c.f. Jahoda 1993). The data were provided 
in rectangular ecliptic coordinates in approximately 
$0.5 \times 0.25$ degree pixels (at ecliptic
equator),  which considerably oversamples the $3 \times
1.5$ degree FWHM beam. 
These data are then corrected for a small systematic 
instrumental change from day $\sim 430$ onwards. 
In this work, we further bin the data into groups of 12 by 12 pixels 
(smaller resolution pixels are strongly correlated due to the 
instrument beam) for all analyses. At the
ecliptic equator the pixel groups are therefore $6^{\circ}\times
3^{\circ}$. Masking (see below) is however performed initially on the
higher resolution data and the final pixel groups contain 
the {\em mean} count rate of all non-zero `sub-pixels' and are 
weighted according to their area. 

It is difficult to unambiguously separate foreground (Galactic) 
from background (extragalactic) information in the HEAO1 X-ray data. 
The total number of resolved foreground and background sources 
($|b|>20^{\circ}$) is small ($\sim 0.01$ deg$^{-2}$) and a 
detailed model of possible large scale Galactic emission is
hard to determine. However, the Galactic 2-10 keV emission model 
of Iwan et al.~(1982) predicts variations of no more than $3\%$ 
of the total flux due to smoothly distributed emission of Galactic 
origin at latitudes $|b| >20^o$.
Studies in the soft bands ($<0.75$ keV) by ROSAT (Snowden 1996) 
indicate that, at these lower energies, the picture is more complicated, 
with Galactic emission at all scales. 

In the present work, as a first step towards removing the
foreground, we construct a `mask' using a list of resolved and
identified Galactic X-ray sources (Piccinotti et al.~1982) and a
$|b|<20^{\circ}$ Galactic Plane mask. Regions of sizes varying from
$\sim 8^{\circ}$ to $12^{\circ}$ diameter are excised around
resolved sources, larger regions are removed around the Large and Small
Magellanic Clouds. A total of $\sim 23$\% of the raw all-sky
flux is removed by this `Galactic' mask.

The removal of bright extragalactic sources is also very important 
in order to control shot-noise in the angular power estimates (LPT97).
We attempt to do this by further masking out all
61 extragalactic sources (AGN's and clusters) in the catalogue of
Piccinotti et al.~(1982), to a flux limit of $S_{cut}=3\times 10^{-11}$ erg
s$^{-1}$ cm$^{-2}$ (2-10 keV).  
An additional $\sim 22$\% of the raw all-sky
flux is removed by this `extragalactic' mask
(the combination of large beam and cautiously generous
source excision 
results in $\sim 50 $ deg$^2$ being removed per source).
The final unmasked area is therefore $\sim 55$\% of the sky 
with an effective redshift of $\sim 0.02$ (approximately the median 
redshift of the Piccinotti et al. sources). 

Finally, the dipolar contribution to the anisotropy due to 
the motion of the observer with respect to the XRB, the 
Compton-Getting (CG) effect, is subtracted from the flux 
(Boldt 1987, Jahoda 1993, LPT97). 
The amplitude of this dipole is estimated from our observed 
motion with respect to the CMB and the observed spectral index
of the hard XRB ($\alpha=0.4$). 
We note that the raw HEAO1 dipole (Galactic sources and plane
removed), due to both the CG effect and large scale structure 
(see LPT97), points in the direction $l \approx 330^o; b\approx 33^o$.
This can be compared with the CMB dipole (in the Local Group frame) 
which points towards $l \approx 268^o; b\approx 27^o$
(based on COBE, Lineweaver et al. 1996). We shall further
discuss the HEAO1 dipole elsewhere (Scharf et al., in preparation).
The HEAO1 data are then expanded in spherical harmonics and 
the harmonic coefficients determined (Scharf et al.~1992, LPT97).

\section{Modeling}

To model the large angular scale fluctuations in the XRB,
we follow the formalism proposed by LPT97
using the following new set of assumptions: 

{\it (i)} X-ray light traces mass, and we assume linear,
epoch-{\it dependent} biasing between the spatial fluctuations
in the X-ray source distribution, $\delta_x$, and those 
in the underlying mass distribution, $ \delta_M$:
$
\delta_x (z) = b_x(z) \delta_M (z).
$
We adopt the following prescription (Fry 1996) for the
time-dependence of the biasing parameter, which we 
parametrize in terms 
of the present-epoch parameter $b_x(0)$:
\beq 
b_x(z)=b_x(0) +z [b_x(0)-1 ]
\eeq
This assumption is somewhat more realistic than
the time-independent bias parameter used by LPT97.
In Fry's model the galaxies are formed at an early epoch
$z_*$ in a biased way, then cluster with time under the influence
of gravity. Note that if $b_x(z_*) =1$ then $b_x(0)=1$.
However, if $b_x(z_*) >1$, biasing decreases with cosmic epoch
(see also Bagla 1998).  

{\it (ii)}
We assume an Einstein-de Sitter cosmology 
($\Omega=1$, $\Lambda=0$),
but we use a phenomenological 
low-density CDM model (with shape parameter $\Gamma=0.2$) 
to represent 
the present-day power-spectrum $P(k)\equiv {\sigma_8}^2 {\bar P}(k)$, 
where $\sigma_8$ is 
the present-epoch 
normalization of the mass fluctuations in 8 $h^{-1}$ Mpc spheres. 
In this case the 
mass  power-spectrum evolves according to linear 
theory as $P (k ,z)\propto (1+z)^{-2}$. 
For the X-ray light fluctuations, $\delta_x({\bf k},z)$, 
the above assumptions translate into:
\beq 
\langle \delta_x({\bf k}) \delta_x^*({\bf k'})  \rangle (z)  =  
(2 \pi)^3 \sigma_8^2 ~b_x^2~(z)~ {\bar P}(k)(1+z)^{-2}
\delta^{(3)}({\bf k} - {\bf k'}), 
\eeq
where $\delta^{(3)}$ is the three-dimensional delta-function.

{\it (iii)} The X-ray intensity observed in the 2-10 keV energy
band originates from the integrated emission of discrete 
X-ray sources out to some high redshift $z_{max}$.
We describe this population by its local luminosity
function $\phi_x(L)$ and spectral index $\alpha$,
and assume simple power-law evolution both in luminosity: 
$L(z)\propto (1+z)^e$, 
and in number density:  
$\phi(L,z)\propto (1+z)^d$.
The local X-ray light density is:
\beq 
\rho_0 = \int_0^{\infty} L\phi_x(L){\rm d}L,
\eeq
and the X-ray light density at redshift $z$ 
{\it observed} in the 2-10 keV energy range is: 
\beq 
\rho_x(z)= \rho_0 (1+z)^q 
\eeq 
where $q=d+e-\alpha+1$.

\medskip

We use the above assumptions to predict the ensemble average 
of the spherical harmonic coefficients in the XRB.
The total predicted signal results in a large scale structure 
component, reflecting the underlying mass distribution,
and a shot noise component due to the discreteness 
of the sources (as opposed to the continuous mass distribution):
\beq
\langle |a_l^m|^2 \rangle_{model} =
\langle |a_l^m|^2 \rangle_{LSS} +\langle |a_l^m|^2 \rangle_{SN}.
\eeq
The shot noise term is:
\beq
\langle |a_l^m|^2 \rangle_{SN} 
= {1 \over 4 \pi} \sum_{sources}  S_i^2
=\int_0^{S_{cut}} S^2 N(S) {\rm d}S, 
\eeq
where $N(S)$ is the differential number-flux relation 
of the X-ray sources. Bright sources (brighter than a suitable
flux cutoff $S_{cut}$) must be removed to reduce the shot noise. 
In turn, removing sources, albeit few and nearby,
will also reduce the large scale structure signal.
However, as we demonstrate below, the shot noise decreases faster
than the signal as more and more sources are removed. 
In other words, the large scale structure signal-to-noise
increases when lowering the flux cutoff.

The large scale structure component can be written as 
an integral over the power spectrum (LPT97):
\beq
\langle |a_l^m|^2 \rangle_{LSS}  =  
{(r_H~ \rho_0)^2 \over (2 \pi)^3}
\int  k^2 {\bar P}(k) |\Psi_l(k)|^2 {\rm d}k,
\eeq   
where $r_H= c/H_0$ is the Hubble radius and
the window function $\Psi_l$ contains the model
parameters:
\beq
\Psi_l(k) = \int_0^{z_{max}} \sigma_8 b_x(z)
(1+z)^{q - 9/2}  j_l(k r_c) W_{cut}(z) {\rm d}z~.
\eeq
The function $W_{cut}(z)$ accounts for the removal of sources
brighter than $S_{cut}$:
\beq 
W_{cut}(z)= {1\over \rho_0} \int_0^{L_{cut}(z)} L\phi_x(L){\rm d}L, 
\eeq
where:
\beq
L_{cut}(z)=4\pi r_c^2(z)S_{cut}(1+z)^{\alpha+1-e}
\eeq
and $r_c(z)$ is the comoving radial distance.
For the monopole ($l=0$), we recover the `Olbers integral': 
$ A_0= \langle |a_0^0|^2 \rangle^{1/2}_{LLS}= \bar I \sqrt {4 \pi}$, where 
$\bar I$ is the mean total intensity of the XRB. In a flat universe
and for $q\ne2.5$, Eq.~4 implies: 
\beq
{\bar I} 
= { \rho_{0} r_H \over 4 \pi} \times
 {(1+z_{max})^{q-2.5} -1 \over q-2.5} .
\eeq 
The higher order multipoles
characterize the spatial fluctuations of the XRB on angular scales
$\sim\pi/l$.

In order to compare model expectations with HEAO1 observations, 
we further convolve our predictions 
with the foreground masks described above:
\beq
\langle |c_l^m|^2 \rangle =\sum_{l' m'} |W_{ll'}^{mm'}|^2
\langle |a_{l'm'}|^2 \rangle,
\eeq
where the $W_{ll'}^{mm'}$ tensor models the mask
(Peebles 1980, Scharf et al.~1992, Baleisis et al.~1998).
Finally, the masked harmonics and shot noise are 
normalized over the monopole. We use the following notation: 
\beq
C_{SN} = {\langle |c_l^m|^2 \rangle_{SN}^{1/2} \over  A_0} 
\eeq
for the shot noise, and for the full signal:
\beq
C_{l}= {(\langle |c_l^m|^2 \rangle_{LSS}+ 
\langle |c_l^m|^2 \rangle_{SN})^{1/2} \over A_0}.
\eeq

\section{Constraints on model parameters}

The local luminosity function in the 2-10 keV energy band 
can be fitted by a double power-law function between
$\sim 10^{42}$ and $10^{48}~h^{-2}~{\rm  ergs~s^{-1}}$ 
(Grossan 1997, Boyle et al.~1998).
The integrated emission of local sources in this range of luminosity 
is: $ \rho_0 \approx  10^{39}~ h~ {\rm ergs~s^{-1}Mpc^3}$.
The total intensity of the 2-10 keV XRB is
${\bar I}= 5.2 \times 10^{-8} {\rm ~ergs~s^{-1}~cm^2~sr^{-1}}$,
and its spectral index is $\alpha=0.4$ (Boldt 1987). 

Boyle et al.~(1998) find evidence for strong cosmological
evolution matching a `pure' luminosity evolution model: 
$L_x \propto (1+z)^e$ with $e \approx 2$ out to a redshift 
of $\sim 2$, followed by a declining phase. 
This scenario would have to hold to $z_{max}\approx 6.4$ 
in order to account for the total XRB intensity (Eq.~11), 
and thus requires other processes or populations whose X-ray 
emission would add to that currently observed. 
On the other hand, Hasinger (1998) argues that 
strong {\it number} density evolution: 
$\phi(L,z)\propto (1+z)^d$ with $d \approx 4$, 
provides a better fit to the ROSAT deep sky survey data, 
implying that the whole XRB intensity should
be accounted for by $z_{max}\approx 1.3$. 
New results from the Hamburg/ESO survey show that 
QSO's keep evolving strongly to $z\sim 3$ and that none of
the above simple parameterizations is an acceptable representation
of the data (Wisotzki, private communication). 
For simplicity however, we shall use the following two toy models
to bracket more realistic X-ray source evolution scenarios:
on the one hand, a `pure' luminosity evolution model with
$q=e-\alpha+1= 2.6$ (see Eq.~4) and $z_{max}=6.4$;
on the other hand, a `pure' density evolution model with
$q=d-\alpha+1= 4.6$ and $z_{max}=1.3$.


We compute the differential number counts relation for both models. 
Both are in good agreement with the Euclidean curve,
$N(S) \propto S^{-2.5}$, derived from ASCA deep sky 
observations to $S \sim 5\times 10^{-14}{\rm ergs~s^{-1}cm^{-2}}$ 
(e.g. Georgantopoulos et al.~1997).
At fainter fluxes both predicted log$N$-log$S$ relations slightly bend 
down to $S\sim 5\times 10^{-16}{\rm ergs~s^{-1}cm^{-2}}$, 
at which flux
the total intensity of the XRB is accounted for. From these number
counts, we derive the shot noise level as a function of 
flux cutoff (Eq.~6). 

\section{Results}

Figure 1 shows the normalized HEAO1 XRB 
harmonics measured through the `Galactic' mask 
(upper panel) and 
through the full, foreground removed mask 
(lower panel) respectively. 
The lower shot noise from bright source removal is immediately 
apparent as a lowering in the overall harmonic amplitude.
The various lines represent our model predictions for the shot 
noise and large scale structure signal, as described below.

Both evolution scenarios yield similar shot noise values within 5\%. 
Masking induces the otherwise constant shot noise to decrease 
slightly (by less than 10\%) towards the high $l$'s. 
As the difference between the two evolution scenarios 
and the gradient due to masking are negligible, 
we have plotted the mean shot
noise value as one horizontal line on both panels in Fig.~1:
for $S_{cut}=3\times 10^{-10} {\rm ergs~s^{-1}~cm^{-2}}$
(i.e.~Galactic sources removed),
$C_{SN}\approx 1.1\times 10^{-3}$;
for $S_{cut}=3\times 10^{-11} {\rm ~ergs~s^{-1}cm^{-2}}$
(i.e. the flux limit of the Piccinotti et al.~1982 catalogue),
$C_{SN}\approx 5.2\times 10^{-4}$.
The predicted shot noise levels (masked and normalized) 
are in very good agreement with the flattening of the 
measured signal in both cases.
We verified this by Maximum Likelihood analysis
over the harmonic range $10 \leq l \leq 20$, ignoring the 
clustering term and leaving the shot-noise level as 
a free parameter. We find that the derived shot-noise 
for both masks is within 10 \% of the one predicted from 
the counts. We also attempted a Maximum Likelihood over the range
$1 \leq l \leq 20 $ with 2 free parameters:
$b_x(0)$ and the shot noise level $C_{SN}$, but the 
2 parameters are strongly coupled.  
 As another independent measure of the shot-noise level
 we have generated a `noise' map, randomly drawing fluxes out of the real
 flux distribution using the $S_{cut}=3\times 10^{-11}$ data (i.e. with
 both Galactic and extragalactic sources from the Piccinotti et al.~catalogue
 removed; The expected CG effect is also removed, as described previously).
 The noise map is then masked as in the data and the $C_l$'s determined.
 The `1-$\sigma$' errors on the mean over 100
 realisations is in excellent agreement with our shot noise 
 estimate independently derived from the source counts (Eq~6).

The first harmonics, $l=1-3$, are well above the shot-noise 
level on both panels, but higher order harmonics are just over 
1-$\sigma$ away from the `noise' estimate.  
(Note that the harmonic measurements are not independent, 
due to `cross talk' introduced by the mask.)
Although contamination from Galactic emission or masking may 
be non-negligible, it is nevertheless encouraging that the 
shape of the harmonic spectrum over all $l's$ is 
qualitatively in agreement with the prediction of
an extragalactic clustering signal.

All models for the spherical harmonic spectrum plotted in Fig.~1
assume a low density CDM model with shape parameter $\Gamma=0.2$ 
and normalization $\sigma_8=1.0$. The present-epoch bias 
parameter $b_x(0)$ (Eq.~1) was left as a free parameter and
its optimal value derived from Maximum Likelihood over the range 
$1 \leq l \leq 10$ (neglecting the mask, cf. Scharf et al. 1992).
Both evolution scenarios yield the same best fit values:
$b_x(0)=1.6$ for the Galactic mask, and $b_x(0)=1.0$ for the full mask. 
To illustrate our estimate range,
predictions are plotted for these 2 values on $both$
panels (Galactic mask and full mask): upper lines are
for $b_x(0)=1.6$ and lower lines for $b_x(0)=1.0$. 
The dotted lines represent the density evolution scenario
($q=4.6$) and the long-dashed lines
show the luminosity evolution scenario ($q=2.6$). 
Assuming a standard CDM power spectrum yield $b_x(0) = 1.8$ and $1.2$
for Galactic and full masks, respectively. This is not surprising, 
as low density CDM has more power on large scales than standard CDM.

For most of the models considered above, 
the reduced $\chi^2$ is near unity,
suggesting acceptable fits.
However, the measured multipoles do show more curvature 
as a function of $l$ than our models predict.
This may be explained by a number of reasons:
the low order multipoles measured in the XRB may result 
from local (Galactic?) structures unaccounted for by the masks;
source clustering evolution may be significantly 
stronger than the linear theory assumption we have made;
or else, the evolution parameters we have used 
for the X-ray source population are overestimated,
at least on part of the redshift range. 
Note also that the $b_x(0)=1$ models, which correspond to 
constant biasing (see Eq.~1), are flatter than 
the epoch-dependent biasing models. Therefore we expect
stronger bias evolution to improve the fit to 
the data.






Figure 2 shows the signal-to-noise as a function of
$l$, assuming  $b_x(0) =1$ and $\sigma_8=1$ (for the purpose of 
illustration). In order to compare the above models with predictions
at fainter flux limits, we do not use the existing masks.
In the lower two panels, we show the signal-to-noise expected
if sources brighter than $S_{cut}=3\times 10^{-12}$ and
$3\times 10^{-13}{\rm ~ergs~s^{-1}cm^{-2}}$ respectively
(i.e. 1 and 2 magnitudes lower than the present data) 
could be removed from the X-ray all-sky survey. We predict
the signal-to-noise to increase as 
$S_{cut}$ decreases. The multipoles are also expected to
be detectable above shot noise for an increasing range
of $l$. As the present results suggest $b_x(0) \ge 1$,
the signal-to-noise ratios plotted here may be taken as lower limits.
Luminosity evolution and density evolution become increasingly
distinct as we remove fainter and fainter sources.
If sources evolve in luminosity, a given flux cutoff will
span a larger redshift range than if they don't or if they
only evolve in number, and therefore
a larger volume of space will be excluded from the analysis.
Hence a weaker signal-to-noise in the case of luminosity 
evolution than in the case of density evolution.
We conclude that an X-ray all-sky survey in the hard band
(to minimize Galactic contamination)  
resolving sources only one magnitude fainter than HEAO1, 
is likely to reveal large scale fluctuations
in the background to significantly higher
order than the current data.

Figure 3 shows the amplitude of rms fluctuations, 
$({ {\delta \rho} \over {\rho} })^2 \sim k^3 P(k) $,
derived at the 
effective scale $k^{-1} \sim 600 h^{-1}$ Mpc probed by the XRB quadrupole 
(cf. LPT97 Figure 1). 
For Galactic mask and either evolution models we find 
$\sigma_8 b_x(0) \sim 1.8 $ and $1.6$ for standard and low density 
CDM models, respectively
(marked by the top and
bottom crosses). 
The fractional error on the XRB amplitudes (due to the
shot-noise of the X-ray sources) is about 30\%.
We see that the observed fluctuations in the XRB
are roughly as expected from interpolating between the
local galaxy surveys and the COBE CMB experiment.
The rms fluctuations 
${ {\delta \rho} \over {\rho} }$
on a scale of $\sim 600 h^{-1}$Mpc 
are less than 0.2 \%.
Our estimate of the fluctuations
amplitude derived  from HEAO1 is used elsewhere 
(Wu, Lahav \& Rees 1998) to show that the fractal 
dimension of the universe is very close to 3 (to within $10^{-4}$) 
on the very large scales.
This XRB measurement strongly supports the validity 
of the Cosmological Principle (Peebles 1993).

\section{Discussion}

We report on the possible detection of low-order spherical harmonic
modes in the HEAO1 XRB map. Although one must be cautious about the 
interpretation of the signal as being purely extragalactic, 
it is encouraging that the measurements are in agreement with 
{\it a priori} predictions. We find that the XRB fluctuations 
on scales of a few hundred Mpcs are consistent with the 
result of interpolating between fluctuations derived 
from local galaxy surveys and those derived from the 
COBE CMB measurements.

%
%
Various models for the matter density fluctuations and 
the evolution of X-ray sources yield 
present-epoch biasing factor of typically $b_x(0) \sim 1-2 $.  
The present analysis allows for epoch-dependent biasing, 
which seems to give a more reasonable fit than a time-independent 
biasing model, although both schemes yield similar values of $b_x(0)$.
Regarding models of density fluctuations, 
as expected the low density CDM  model requires lower $b_x(0)$
than standard CDM, which has less power on large scales.
We note that our values for the local bias factor $b_x(0)$ are 
smaller than those derived from the dipole anisotropy 
of the local AGN distribution (Miyaji 1994) 
and from HEAO1 assuming epoch-independent biasing 
(Boughn et al. 1998). On the other hand, our results 
are in rough agreement with Carrera, Fabian \& Barcons
(1997) who also find small values of $b_x(0)$,
although using quite different techniques and data sets,
and assuming epoch-independent biasing.

We predict that an X-ray all-sky survey 
resolving sources a factor of 10 fainter than HEAO1, 
may allow us to measure large scale fluctuations
in the XRB to order $l\sim 20$ ($\theta \sim \pi/20$).
The present data cannot be used with lower flux thresholds as our
method of eliminating sources also reduces sky coverage.  
There are two experimental approaches which can allow a
similar analysis to be done while employing a lower flux threshold,
thus reducing the shot noise, and allowing a significant measurement 
of large scale structure over a larger range of $l$ values.

Barcons et al. (1997) propose an experimental concept that maps the X-ray
sky with a collimated proportional counter, substantially similar to the
A2 experiment, but with a smaller field of view.  Such an experiment can
mask individual sources with a smaller penalty in terms of sky coverage 
(and signal) and can therefore mask a larger number, reaching a fainter
limiting flux. But it cannot identify sources an order of magnitude fainter 
than our flux threshold by itself, and therefore would rely on an externally 
generated catalogue such as that produced by the
ABRIXAS survey (Tr\"umper et al.~1998). The advantages of this approach are
the relatively small size and simplicity of the experiment.
The combination of ABRIXAS with the experiment proposed by 
Barcons et al.~(1997)
can eliminate sources down to a flux threshold 
$\sim 1 \times 10^{-12} {\rm erg}\,{\rm sec}^{-1}\,{\rm cm}^{-2}$.  
For the expected density of sources (0.3 per square degree)
and solid angle of the beam for this experiment (1 square degree), 
the fraction of
the sky that will be masked is comparable to the analysis presented here. 
This experiment represents the limiting capability of a non focussing 
mission.  
The time required to obtain a certain precision per pixel scales inversely 
with solid angle of the beam, and the fraction of the sky which is masked 
out remains large.

Removing sources at fainter thresholds requires an imaging experiment capable
of identifying, and excluding, faint sources without removing an entire
square degree of sky coverage.
Due to the relative inefficiency of X-ray telescopes (the effective
area is typically $\le$ 25\% of the geometric collecting area above 2 keV
(Serlemitsos \& Soong 1996)) a large collecting area 
(i.e. many telescopes) is
required to obtain the same precision in the measurement of surface 
brightness.
A collection of imaging telescopes capable of measuring surface brightness
to 1\% per square degree requires nearly 3 times the geometric collecting
area of the proportional counter experiment, but is plausibly within the
constraints of NASA Medium Explorer mission (Jahoda 1998).  An additional
advantage of an imaging experiment is the simultaneous production of 
a catalogue
of sources, useful in their own right as tracers of large scale structure.
An experiment capable of identifying sources as faint as 
$3 \times 10^{-13} ~{\rm erg~cm^{-2}~sec^{-1}}$ 
in the 2-10 keV band would generate
an all-sky catalogue with $\ge 10^5$ hard X-ray selected sources.

However `old' and not optimally suited for this analysis,
the current data have allowed us to demonstrate that
future X-ray missions stand a good chance of revealing 
significant structure in the matter distribution. 
Not only are we likely to finally understand the long
sought for sources of the hard X-ray background in the coming 
years, but we may also be able to get a strong hold on the 
underlying matter distribution in an otherwise little 
explored range of scales.

\acknowledgments
The authors thank B. Boyle, M. Rees and K. Wu for
helpful discussions.  We thank the referee, D. Helfand,
for a careful reading and constructive suggestions that have
improved this paper.

%
%
\begin{figure}
\unitlength1cm
\begin{picture}(15,15)
{\epsfxsize=15.cm
\epsfbox{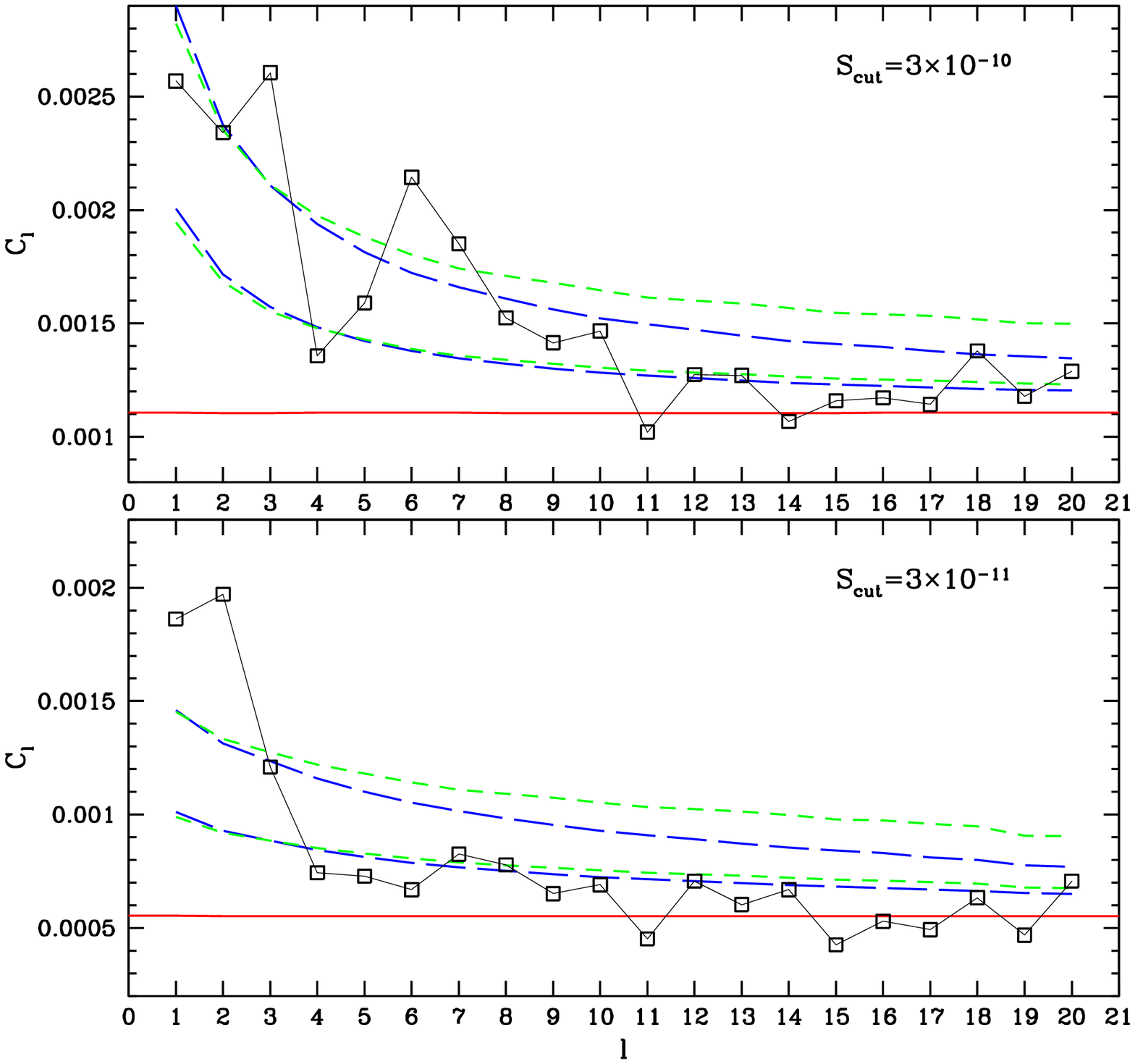}}   
\end{picture}
\vskip -0.5cm 
\figcaption{The normalized spherical harmonic spectrum of the
HEAO1 XRB (squares). In the upper panel, only the Galactic
component was removed from the data. In the lower panel,
extragalactic X-ray sources from the Piccinotti et al.~1982
catalogue were also removed. 
The corresponding flux cutoffs $S_{cut}$ 
in ${\rm ergs~s^{-1}cm^{-2}}$ and predicted levels of shot noise
(horizontal lines) are indicated in both cases.
The dotted lines are the predictions (Eq.~5) of a 
`pure' density evolution model ($q=4.6$), 
and the long-dashed lines are the predictions of a 
`pure' luminosity evolution model ($q=2.6$). 
A low density CDM power-spectrum with normalization $\sigma_8=1$ 
for the mass fluctuations was assumed. 
Upper lines on both panels are for present-epoch bias parameter 
$b_x(0)=1.6$ and lower lines for $b_x(0)=1.0$.
}
\end{figure}

%
%
\begin{figure}
\unitlength1cm
\begin{picture}(15,15)
{\epsfxsize=15.cm
\epsfbox{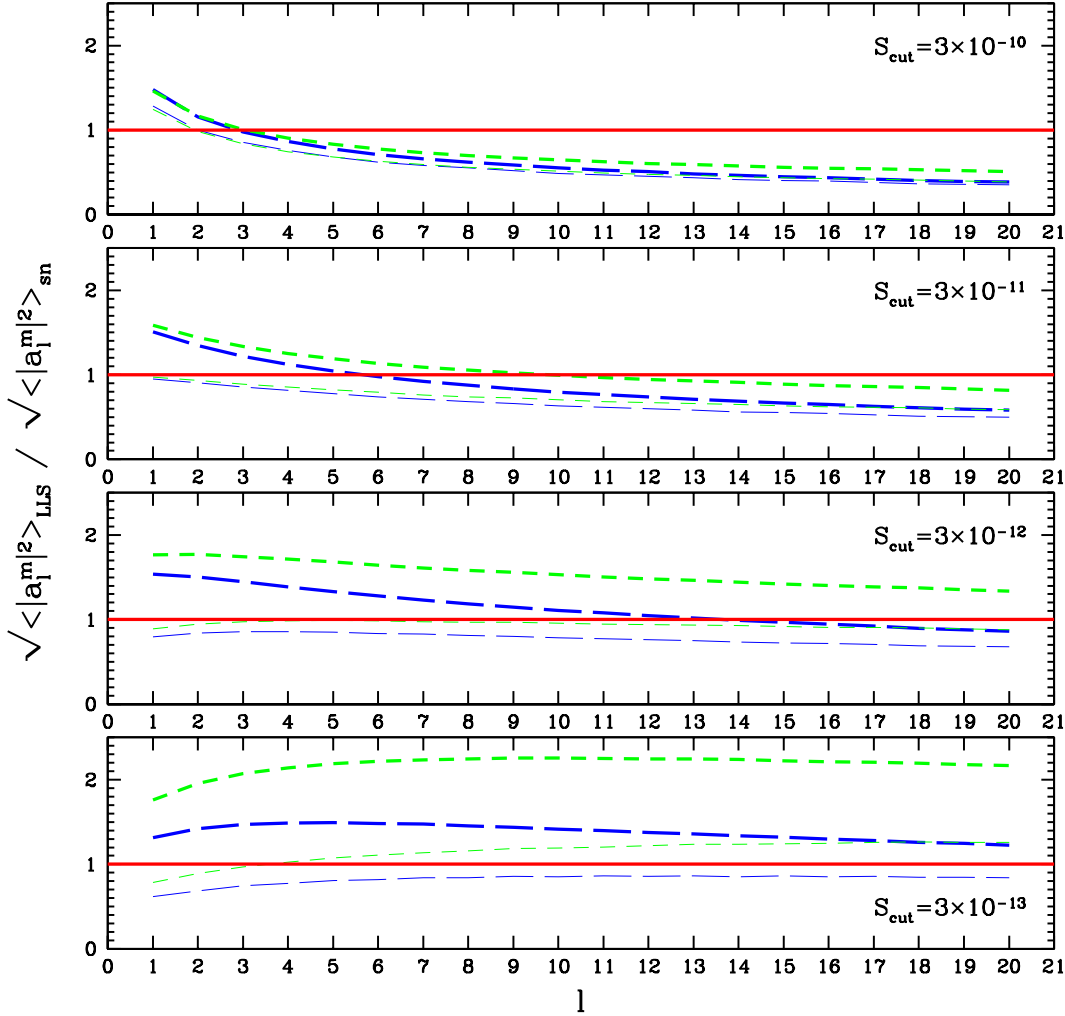}}
\end{picture}
\vskip -0.5cm
\figcaption{Predicted signal-to-noise as a function of $l$
for decreasing flux cutoffs $S_{cut}$ (as indicated on each
panel in ${\rm ergs~s^{-1}cm^{-2}}$). We assumed $b_x(0)=1 $ and 
$\sigma_8$=1. The dotted lines are the predictions of a `pure' 
density evolution model ($q=4.6$), 
and the long-dashed lines are the predictions of a `pure' 
luminosity evolution model ($q=2.6$).
The thick lines assume a low density CDM power spectrum
as described in the text. 
The thinner lines correspond to a standard CDM power-spectrum,
for comparison. The horizontal line (signal/noise $=1$) 
marks the shot-noise level (also the 1-$\sigma$ error bar).
}
\end{figure}

%
%
\begin{figure}
\unitlength1cm
\begin{picture}(15.,15.) 
{\epsfxsize=15.cm
\epsfbox{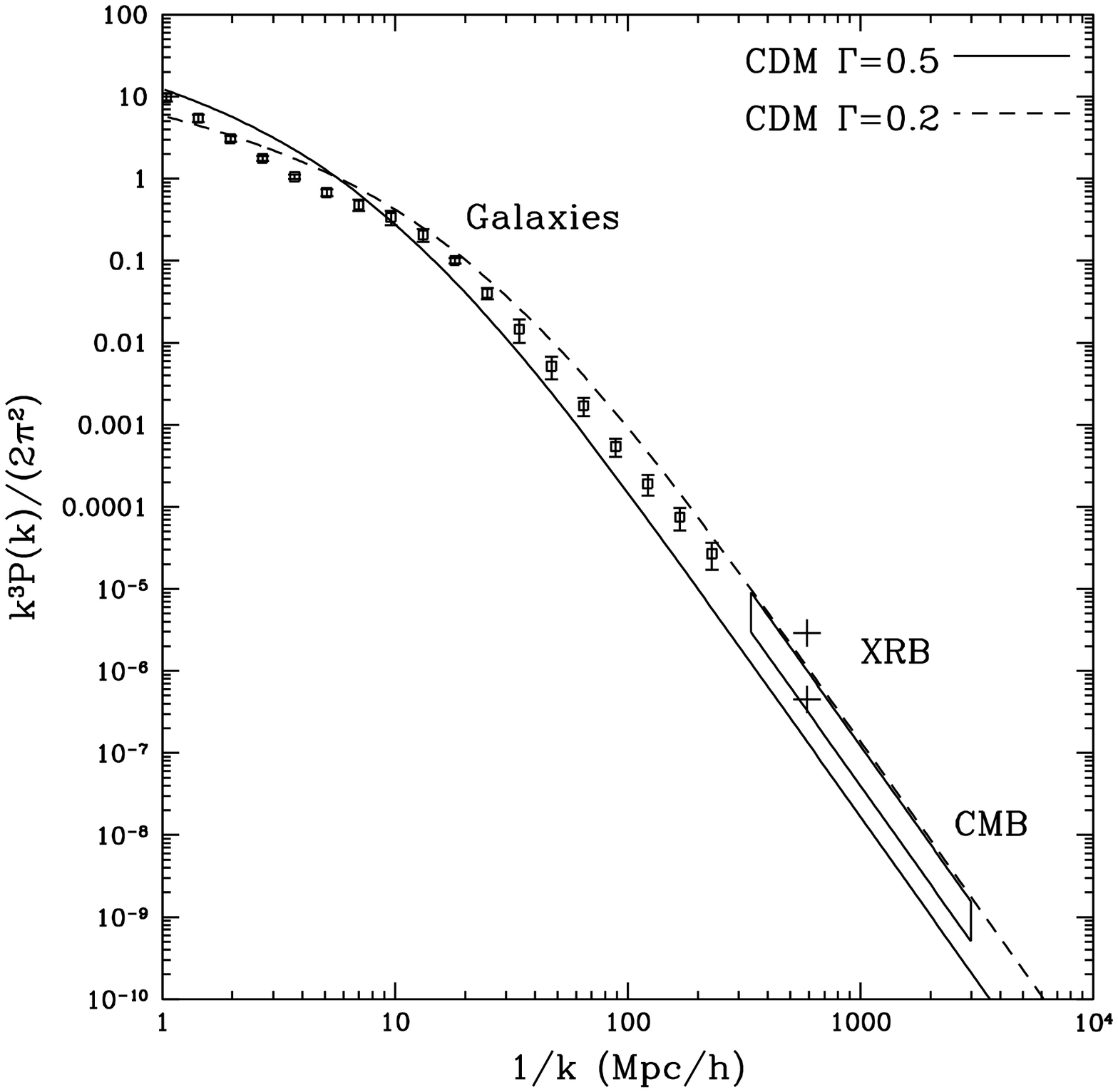}}
\end{picture}
\vskip -0.5cm
\figcaption{
A compilation of rms density fluctuations, 
$({ {\delta \rho} \over {\rho} })^2 \sim k^3 P(k)$,
on different scales 
from various observations: a galaxy survey, the X-ray
Background and Cosmic Microwave Background experiments.  
The crosses represent our present constraints from 
the XRB HEAO1 quadrupole.
The top and bottom crosses are estimates of the amplitude of 
the power-spectrum at $k^{-1} \sim 600 h^{-1}$ Mpc,
assuming CDM power-spectra with shape parameters
$\Gamma=0.2$ and $0.5$ respectively, and an Einstein-de Sitter
universe.  The fractional error on the XRB amplitudes (due to the
shot-noise of the X-ray sources) is about 30\%.
The solid and dashed lines correspond to the standard CDM
power-spectrum (with shape parameter $\Gamma = 0.5$) and a
`low-density' CDM power-spectrum (with $\Gamma=0.2$), respectively,
assuming $\sigma_8=1$ in both cases.
The open squares at small scales are estimates of the
power-spectrum from 3D inversion of the angular APM galaxy
catalogue (Baugh \& Efstathiou 1993, 1994). 
The elongated box at large scales
represent the COBE 4-yr  CMB measurement
(Smoot et al. 1992, Bennett et al. 1996).
The COBE box corresponds to a quadrupole Q=18.0 $\mu K$ for a
Harrison-Zeldovich mass power-spectrum, via the Sachs-Wolfe 
effect, or $\sigma_8 =1.4$ 
for a standard CDM model (Gawiser and Silk 1998).  
}
\end{figure}


\begin{references} 

\reference {} Bagla, J. 1998, MNRAS 297, 251. 
\reference {} Baleisis, A., Lahav, O., Loan, A. \& Wall, J.V. 1998,
              MNRAS, 297, 545
\reference {} Barcons, X., Fabian, A.C., Carrera, F. 1997, 
      	MNRAS 285, 820.
\reference {} Baugh, C. M. \& Efstathiou, G. 1993, MNRAS 265, 145.
\reference {} Baugh, C. M. \& Efstathiou, G. 1994, MNRAS 267, 323.
\reference {} Bennett, C. L. et al.~1996, ApJ 464, L1.
\reference {} Boldt, E. A. 1987, Phys. Reports 146, 215. 
\reference {} Boughn, S., Crittenden, R. \& Turok, N. 1998, 
	New Astronomy, vol. 3, no. 5, p. 275.
\reference {}  Boyle, B.J., Georgantopoulos, I., Blair, A.J., Stewart, G.C.,
	Griffiths, R.E., Shanks, T., Gunn, K.F., Almaini, O. 1998, MNRAS 296, 1. 
\reference {} Carrera, F., Barcons, X., Fabian, A.C. 1997, MNRAS 285, 820.
\reference {} Comastri, A., Setti, G., Zamorani, G., Hasinger, G.
	1995, A\&A 296, 1. 
\reference {} Fabian, A. C., Barcons, X. 1992, ARA\&A, 30, 429. 
\reference {} Fry. J. 1996, ApJ, 461, L65.
\reference {} Gawiser, E. \& Silk, J., 1998, Science, 280, 1405
\reference {} Georgantopoulos, I., Stewart, G., Blair, A., 
	Shanks, T., Griffiths, R.E., Boyle, B., Almaini, O., Roche, N.
	1997, MNRAS 291, 203.
\reference {} Grossan, B.A. 1992, Ph.D. thesis, MIT.
\reference {} Hasinger, G. 1998, Astron. Nachr 319, 37.
\reference {} Hasinger, G., Burg, R., Giacconi, R., Schmidt, M., 
	Tr\"umper, J., Zamorani, G. 1998, A\&A, 329, 482.
\reference {} Iwan, D., Shafer, R.A., Marshall, F.E., Boldt, E.A.,
	Mushotzky, R.F., Stottlemyer, A. 1982, ApJ, 260, 111.
\reference {} Jahoda, K. 1993, Adv. Space Res. 13, (12) 231.
\reference {} Jahoda, K. 1998, Astron. Nachrichten, 319, 129.
\reference {} Lahav, O., Piran, T., Treyer, M. A. 1997, 
	MNRAS, 284, 499 (LPT97).
\reference {} Lineweaver, C., Tenorio, L., Smoot, G., Keegstra, P., Banday,
	A., Lubin, P. 1996, ApJ, 470, 38.  
\reference {} McHardy, I. et al.~1998, Astron. Nachrichten, 319, 51.
\reference {} Miyaji, T. 1994. PhD thesis, University of Maryland.
\reference {} Peebles, P.J.E. 1980, Large Scale Structure of the 
	Universe (Princeton University Press).
\reference {} Peebles, P.J.E. 1993, Principles of Physical Cosmology
        (Princeton University Press).
\reference {} Piccinotti, G., Mushotzky, R. F., 
	Boldt, E. A., Holt, S. S., Marshall, F. E., Serlemitsos, 
	P. J., Shafer, R. A. 1982, ApJ, 253, 485.
\reference {} Scharf, C. A., Hoffman, Y., Lahav, O., 
	Lynden-Bell, D. 1992, MNRAS, 256, 229.
\reference {} Serlemitsos, P.J., \& Soong, Y. 1996, Astrophysics
	and Space Science, 239, 177.
\reference {} Snowden, S. L. 1996, in R\o ntgenstrahlung
 	from the Universe, eds. Zimmermann, H.U., Tr\"umper, J., and 
	Yorke, H., MPE Report 263, 299.
\reference {} Smoot G., et al.,  1992, ApJ L, 396, L1.
\reference {} Tr\"umper, J., Hasinger, G., \& Staubert, R. 1998, 
	Astron. Nachrichten, 319, 113. 
\reference {} Wu, K.K.S., Lahav, O. \& Rees, M.J., 
              1998, submitted to Nature (astro-ph/9804062)

\end{references}
\end{document}